\documentstyle[twoside,fleqn,espcrc2,epsf]{article}

\newcommand{\AmS}{{\protect\the\textfont2
  A\kern-.1667em\lower.5ex\hbox{M}\kern-.125emS}}

\title{
\vspace{-5.0cm}
\begin{flushright}
{\normalsize DOE/ER/40561-274-INT96-00-135}\\
\vspace{-0.3cm}
{\normalsize UW/PT-96-14}\\
\vspace{-0.3cm}
{\normalsize RU-96-67}\\
\end{flushright}
\vspace*{2.5cm}
Overlap for 2D chiral U(1) models}

\author{Rajamani Narayanan
        \address{Institute for Nuclear Theory, Box 351550,
	University of Washington, Seattle, WA 98195-1550}
        \thanks{Research supported in part by the DOE under grants \#
	DE-FG06-91ER40614 and \# DE-FG06-90ER40561}
        and 
        Herbert Neuberger
	\address{Department of Physics and Astronomy, 
	Rutgers University, Piscataway, NJ 08855-0849}
        \thanks{Research supported in part by the DOE under grant \#
	DE-FG05-90ER40559}
	}
       
\begin{document}
\begin{abstract}
The overlap formulation is applied to an anomaly free combination of
chiral fermions coupled to U(1) gauge fields on a 2D torus. Evidence
is presented that gauge averaging the overlap phases in these models
produces correct continuum results.
\end{abstract}

\maketitle

\section{Introduction}

Two dimensional chiral gauge theories provide
a convenient testing ground for non-perturbative regularizations since
they are much simpler to simulate numerically than four dimensional 
ones.
As in four dimensions, there are anomalies in two dimensions that
have to be cancelled and anomaly free theories 
can have global charges that
are broken by topologically non-trivial gauge fields.
In this talk, we will discuss a specific anomaly free
abelian chiral gauge theory 
in two dimensions. The non-perturbative regularization we will employ
is the overlap formulation on the lattice \cite{NN95}. 
The abelian chiral gauge theory we consider contains four left
handed Weyl fermions and one right handed Weyl fermion living on
an $l\times l$ torus. All four
left handed fermions have a U(1) charge equal to 1 and the right
handed fermion has a U(1) charge equal to 2. 

The model is expected to be gauge invariant in the continuum 
since we have chosen the charges to cancel the perturbative anomaly.
On the torus the model is sensitive to the boundary conditions. Even
in the absence of an electric field, nonperturbative violations of
gauge invariance under certain singular gauge transformations can occur if
the boundary conditions are not chosen with care \cite{NN96}. 
A choice free of problems
is that the right handed fermions obey anti-periodic boundary conditions 
while the left handed fermions obey:
$$
\psi_{L1}(x+l\hat\mu) = i\psi_{L1}(x),
    \psi_{L3}(x+l\hat\mu) = - is_\mu\psi_{L3}(x),
$$
$$
\psi_{L2}(x+l\hat\mu) = is_\mu\psi_{L2}(x),
   \psi_{L4}(x+l\hat\mu) = -i\psi_{L4}(x).
$$
where $s_1=1$ and $s_2=-1$. 

Although the theory is chiral, 
with the above choice, the fermion determinant becomes 
real for any vector potential. While the fermion determinant is real
in the continuum limit, the lattice overlap is a complex number 
reflecting the absence of exact gauge invariance at finite lattice spacing.
When the gauge fields are integrated over, gauge invariance is restored by 
group averaging. We shall show that gauge integration along a fixed typical
orbit reproduces the continuum result for that orbit.

\section{Formulation on the lattice}
We embed an $L\times L$ lattice in the continuum $l\times l$ torus. 
To the plaquette with corners at 
$n$, $n+\hat\mu$, $n+\hat\nu$ and $n+\hat\mu+\hat\nu$ we associate
an angle $\phi(n)$.
$\phi(n)$ is a discretization of the continuum $\phi(x)$ related to
the electric field by $E(x)=\partial^2 \phi(x)$.
We will restrict
ourselves to the zero topological sector and therefore 
$\sum_n \phi(n)=0$. 
Given $E(x)$ we can solve for $\phi(x)$ where $\phi(x)$ is a periodic function
on the torus with no zero modes.
The parallel transporters are 
$$
U_\mu(n)= g(n+\hat\mu) g^*(n)
            e^{i\epsilon_{\mu\nu}[\phi(n)-\phi(n-\hat\nu)]}
            e^{i{2\pi\over L}h_\mu}.
$$
The $h_\mu$'s are the zero modes of the gauge potential restricted to 
$[-1/2, 1/2)$.
Gauge orbits will be labeled by $E(n)=(\partial^*_1\partial_1 +
\partial^*_2\partial_2)\phi(n)$, $h_1$ and $h_2$.
$g(n)$ is a U(1) valued group variable on the site $n$ and
labels points on the orbit.
$g(n)=1$ corresponds to the gauge field in the Landau gauge.
The Wilson gauge action is
$$
S^w_g= {1\over e^2}
       \sum_n {\rm Re} [ 1- \cos 
((\partial^*_1\partial_1 +
\partial^*_2\partial_2)\phi(n))]
$$
The fermionic path integral is defined on the lattice using the
overlap formalism \cite{NN95} and we refer the reader to \cite{NN96}
for details.

\section {Overlap along gauge orbits}

The overlap formula will not be gauge invariant on the lattice but 
gauge invariance is restored when the lattice spacing goes to zero
while the gauge field is fixed. Thus, the gauge breaking is expected 
to be small.
In our previous work \cite{NN95}
we suggested dealing with the extra gauge breaking terms
by simply averaging over each gauge orbit. If the breaking is 
not too large, and
if anomaly free chiral gauge theories in the continuum
exist also beyond perturbation theory,
the most plausible outcome is that the averaging along the orbit
simply adds some irrelevant local gauge invariant terms to the
rest of the action. For example, a gauge breaking
term in an action for a pure gauge theory that has the form of
a mass term for the gauge bosons, when averaged
over the gauge orbits, induces only effects irrelevant in the
infrared, as long as its coefficient is not too large \cite{FNN}.

In this section we
shall repeatedly start from some configuration that has a typical
gauge invariant content and average over its gauge orbit
by computing the overlap for many gauge transformations of
the original configuration. The overlap enjoys the nice property
that all the gauge breaking is restricted to
its phase \cite{NN95}. 
The result of averaging this phase will be
some complex number, ${\cal Z}$. 
We will first focus on the resulting phase $\Phi$ where
${\cal Z} =
|{\cal Z}|e^{i\Phi}$.
We will
look at the distribution of the overlap phases along a particular orbit.
We will show that this distribution is quite well peaked around zero
phase indicating that the continuum determinant is real. 
The width of the distribution will be a function of the
gauge invariant content of the orbit. 

The simplest background one could imagine is one in which
there is no electric field and all Polyakov loops are trivial.
The lattice overlap on this ``trivial orbit'' ($\phi=0$ and $h_\mu=0$)
for a single chiral fermion was proven to be real
in \cite{NN95}. 
We will generate a typical
electric field configuration by choosing some $eL$ 
in the Wilson action. 
We pick $h_\mu$ independently in the range $[-1/2,1/2)$. 
In figure 1, we plot the distribution of the phases on orbits where $h_\mu=0$.
The x-axis is the phase in units of $\pi$. 
All the four distributions in the plot are well peaked and centered around zero.
As $eL$ gets larger the distributions gets broader. This is because we are
going away from the trivial orbit. The plot was obtained on a $6\times 6$ lattice
by randomly generating a total of 1000 points along the orbit. In figure 2, we
have set $\phi=0$ and studied the distribution for two different values of 
$h_\mu$. The distribution is again peaked around zero. 
\begin{figure}
 \epsfxsize=3.25truein
\centerline{\epsffile{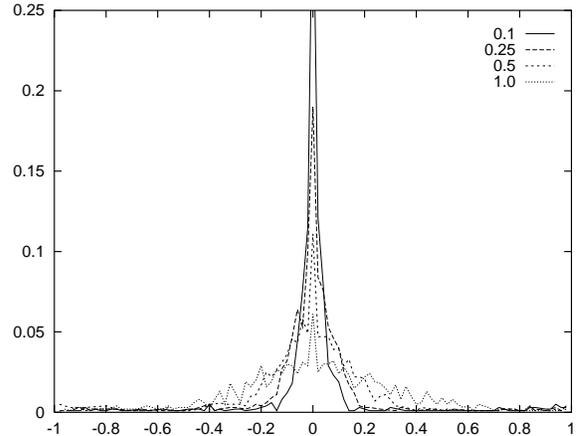}}
\caption{$eL=(0.1,0.25,0.5,1.0)\pi$, $h_1=h_2=0$}
\end{figure} 
\begin{figure}
 \epsfxsize=3.25truein
\centerline{\epsffile{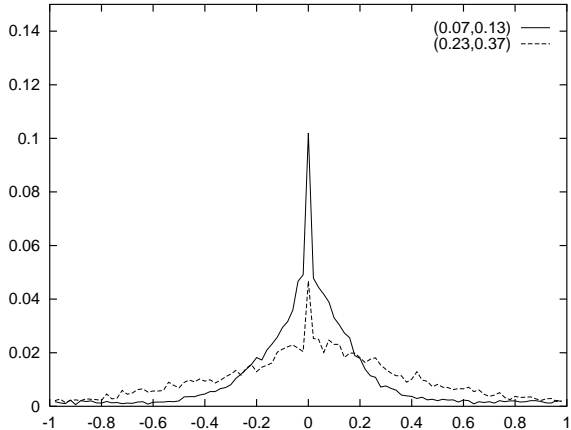}}
\caption{$\phi=0$, $h_1=0.07,h_2=0.13$; $h_1=0.23,h_2=0.37$}
\end{figure} 

Figures 1 and 2 show that
one can perform an integration along the orbit. Integration will induce an
additional term to the effective action. This term will be proportional to the
width of the distributions and we can study it as a function of $\phi$.
The width characterizes the gauge invariant content of the orbit. 
Note that the continuum determinant depends on $\phi$ via a factor
$\exp{-{2\over\pi}\int d^2 x \phi \partial^2 \phi }$ \cite{NN95}.
In two dimensions, a Thirring term can be induced upon regularization since
it is renormalizable. This was found to be the case in a vector theory
\cite{NNV}. 
Such an effect is expected to be present in the chiral model
also since the modulus of the overlap is the same as that of a vector theory. 
The integration over the orbit 
also induces a Thirring type interaction. 
The two Thirring terms modify the coefficient in front of
the $\int  d^2 x \phi \partial^2 \phi $ in the exponent of the overlap:
$${2\over \pi} \rightarrow {2\over \pi+g^2_1} + g^2_2 \equiv {2c\over \pi}$$
$g_1$ is the effect present in the modulus of the overlap and $g_2$
is the effect of integration over the orbit. The combined effect is
denoted by one coefficient $c$. 

We extract the coefficient $c$ by plotting 
the induced term on the lattice (``lattice''), 
obtained by the taking the logarithm of the overlap integrated over
the orbit, as a function of 
${2\over\pi}\int d^2 x \phi \partial^2 \phi$ (``continuum'')
at fixed $el=\pi$. 
The imaginary part of the lattice term, as discussed
previously, is consistent with zero.
In Figure 3 we present a scatter-plot of the lattice term vs. the 
continuum one including several typical $\phi$ configurations and several
lattice spacings. 
Note that the points align quite well along a line with a slope of
about unity. The evident correlation between the values on the two axes
indicates that indeed one can parameterize the induced action by $c$ and
we find that $c\approx 1$ in agreement with the continuum. In particular
the result implies that the ``photon mass'' comes out correctly in this model.

\begin{figure}
 \epsfxsize=3.25truein
\centerline{\epsffile{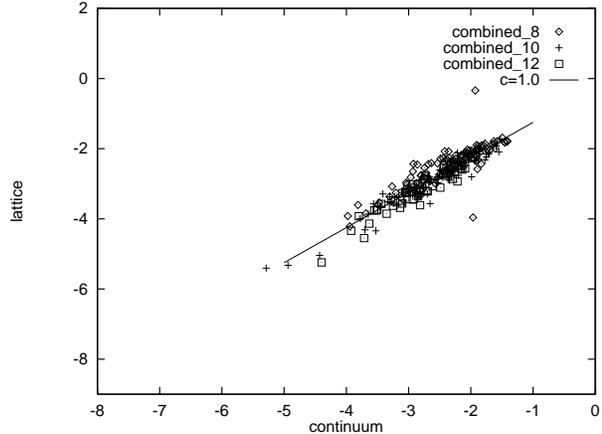}}
\caption{Combined action, $h_1=h_2=0$, $eL=\pi$}
\end{figure}

\end{document}